\documentclass{article}
\usepackage{spconf,amsmath,graphicx}

\usepackage{xcolor,soul,framed} %,caption

\colorlet{shadecolor}{yellow}

\usepackage{array}
\usepackage{url}

\usepackage{cite}
\usepackage{booktabs} 
\usepackage{multirow}
\usepackage{multicol}
\usepackage{amsfonts}
\usepackage{float}

\usepackage{hyperref}
\usepackage{etoolbox,siunitx}
\robustify\bfseries
\usepackage{balance}
\usepackage{comment}
\usepackage{amssymb}% http://ctan.org/pkg/amssymb
\usepackage{pifont}% http://ctan.org/pkg/pifont
\newcommand{\cmark}{\ding{51}}%
\newcommand{\xmark}{\ding{55}}%

\usepackage{scalerel}
\title{Scenario-Aware Audio-Visual TF-GridNet for Target Speech Extraction}
\name{\parbox{\linewidth}{\centering Zexu Pan$^{1}$, Gordon Wichern$^{1}$, Yoshiki Masuyama$^{1,2}$, Fran\c{c}ois G. Germain$^{1}$,\\Sameer Khurana$^{1}$, Chiori Hori$^{1}$, Jonathan Le Roux$^{1}$\thanks{This work was performed while Y.~Masuyama was an intern at MERL.}}}
\address{
  $^{1}$Mitsubishi Electric Research Laboratories (MERL), Cambridge, MA, USA\\
  $^{2}$Tokyo Metropolitan University, Tokyo, Japan} 

\begin{document}
%\ninept
%
\maketitle
\begin{abstract}
Target speech extraction aims to extract, based on a given conditioning cue, a target speech signal that is corrupted by interfering sources, such as noise or competing speakers. Building upon the achievements of the state-of-the-art (SOTA) time-frequency speaker separation model TF-GridNet, we propose AV-GridNet, a visual-grounded variant that incorporates the face recording of a target speaker as a conditioning factor during the extraction process. Recognizing the inherent dissimilarities between speech and noise signals as interfering sources, we also propose SAV-GridNet, a scenario-aware model that identifies the type of interfering scenario first and then applies a dedicated expert model trained specifically for that scenario. Our proposed model achieves SOTA results on the second COG-MHEAR Audio-Visual Speech Enhancement Challenge, outperforming other models by a significant margin, objectively and in a listening test. We also perform an extensive analysis of the results under the two scenarios.
\end{abstract}
\vspace{-2mm}
\section{INTRODUCTION}
\vspace{-2mm}
\label{sec:intro}
Speech, as the most natural form of human communication, effectively delivers rich information about a speaker's emotion, identity, location, or spoken content. Many speech processing algorithms have been developed to extract such information~\cite{pan2020multi, snyder2018x, qian2021multi, wang2022predict,tao2021someone}. These algorithms are often optimized for clean speech signals, while real-world speech signals are typically contaminated by interfering signals such as noise and irrelevant speakers, which is exemplified as the ``cocktail party problem''~\cite{bronkhorst2000cocktail}. Therefore, it is often beneficial to incorporate a pre-processing step to extract the speech signal of interest, a task commonly known as target speech extraction~\cite{zmolikova2023neural}.

Target speech extraction algorithms are usually conditioned on an auxiliary reference or cue, to distinguish the target speaker in the case of overlapping speakers. A widely studied case is that of a pre-recorded speech utterance being used as the reference, where the network extracts the speech that sounds similar to the speech signal reference~\cite{Chenglin2020spex,wang2019voicefilter,spex_plus2020,he2020speakerfilter,xiao2019single,shi2020speaker,delcroix2020improving}. A drawback of such an approach is that pre-enrollment of each target speaker is needed, which is cumbersome or even unfeasible in some circumstances. % for strangers. 

Human attention is known to be multi-modal~\cite{smith2005development}, involving various sensory stimuli that are processed interactively as described by the reentry theory~\cite{edelman1987neural}. Notably, studies have shown that watching a speaker significantly improves speech comprehension in a challenging ``cocktail party'' scenario~\cite{ma2009lip,golumbic2013visual,crosse2016eye}. Motivated by these studies and the robustness of visual cues against acoustic noise, there have been various attempts to condition the target speech algorithm on visual signals. For example, the FaceFilter model explores the face-voice correspondence using a single face image~\cite{chung2020facefilter}, the reentry model explores the speech-lip synchronization using a lip recording~\cite{pan2021reentry}, and the SEG model explores the speech-gesture association using an upper-body recording~\cite{pan2022seg}.

Among visual cues, face recordings are generally understood to be the most effective and are the most often used, as visemes provide the exact places of articulation~\cite{michelsanti2021overview,afouras2018conversation,wu2019time,wu2022time,li2020deep,li23ja_interspeech,tavcse2022,tan2020audio,lu2019audio,morrone2019face}. Researchers have explored the use of face recordings in various target speech networks, such as frequency-domain bidirectional long short-term memory (BLSTM) networks~\cite{ephrat2018looking,ochiai2019multimodal}, time-domain temporal convolutional networks (TCN)~\cite{luo2019conv,wu2022time,pan2020muse,pan2021reentry,pan2023imaginenet}, or dual-path recurrent neural networks (DPRNN)~\cite{luo2020dual,usev21}. In this work, motivated by the recent success of the time-frequency-domain speaker separation model TF-GridNet, we propose to condition it on face recordings for audio-visual target speech extraction, referring to this model as AV-GridNet.

While a target speaker extraction algorithm can work independently to the type of interfering signals, we contend that, the characteristics of speech and noise being very different, it may be more advantageous to individually optimize a model for each interference scenario, as a model separating speech from speech is likely to more heavily rely on the intrinsic structure of speech signals, while a model separating speech from noise is likely to more heavily rely on the differences between the characteristics of speech and noise.

In this paper, on top of a universal AV-GridNet that processes the mixture speech signal irrespective of the type of interfering signals, we thus propose a scenario-aware AV-GridNet model,  SAV-GridNet, that explicitly integrates the different interfering scenarios. SAV-GridNet is a cascaded model that first identifies the type of interfering signals with a classifier model, and then applies a dedicated expert AV-GridNet model that is trained specifically for that scenario. To validate the effectiveness of our proposed networks, we participated in the second COG-MHEAR Audio-Visual Speech Enhancement Challenge~\cite{blanco2023avse}, and achieved state-of-the-art (SOTA) results in terms of objective measures such as perceptual evaluation of speech quality (PESQ), short-time objective intelligibility (STOI), and scale-invariant signal-to-distortion ratio (SI-SDR), in addition to word intelligibility in a listening test, outperforming the other teams and the baseline~\cite{pan2022hybrid} by a significant margin. We also perform an extensive analysis of the results under the two scenarios.

\vspace{-2mm}
\section{METHODOLOGY}
\vspace{-2mm}
\subsection{Related work: TF-GridNet}
\vspace{-2mm}

Our proposed system is built upon the SOTA time-frequency-domain model called TF-GridNet, which has demonstrated promising results in various tasks including speech separation~\cite{wang2023tf} and multi-channel audio-only target speech extraction~\cite{Cornell2022}.
TF-GridNet directly estimates the real and imaginary components of the target speech signal from those of the mixture speech.
It encodes a speech signal into TF representations of dimension $T \!\times\! F \!\times\! D$ with a short-time Fourier transform (STFT) followed by a 2-dimensional convolution (Conv2D) and a layer normalization (LN) operation. Then,
$B$ repetitions of GridNet blocks are applied to refine the TF representations.
Finally, the real and imaginary components of the target speech are obtained by applying 2-dimensional deconvolution to the output of the final GridNet block, followed by an inverse STFT (iSTFT) operation to output the separated speech signals.

Each GridNet block consists of three successive modules:
1) An intra-frame spectral module that views the TF representation as $T$ separate sequences of $D$-dimensional embeddings, each sequence having length $F$, and applies a BLSTM layer with $H$ units and a 1-dimensional deconvolution layer with kernel size $I$ and stride $J$;
2) A sub-band temporal module that views the TF representation as $F$ separate sequences of $D$-dimensional embeddings, each sequence having length $T$, and performs a similar procedure as in the intra-frame spectral module;
3) A full-band self-attention module that first reshapes the TF representation into a single sequence of length $T$ with $F\!\times\! D$ channels, and applies a multi-head self-attention operation with $L$ heads.

\vspace{-2mm}
\subsection{Proposed AV-GridNet}
\begin{figure}
  \centering
  \includegraphics[width=0.8\linewidth]{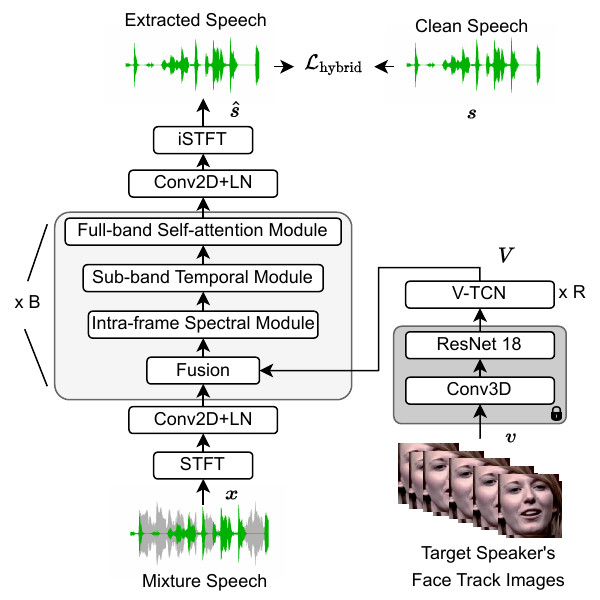}
  \vspace{-4mm}
  \caption{Our AV-GridNet model extracts the target speech conditioned on the target's face recording.}
  \label{fig:av-gridnet}
  \vspace{-4mm}
\end{figure}

\vspace{-2mm}
Instead of separating all speakers into individual streams, we propose a modified version of TF-GridNet for audio-visual target speaker extraction called AV-GridNet. AV-GridNet is conditioned on the face recording $v$ of the target speaker, and it extracts only the corresponding target speech $\hat{s} $ from the mixture speech signal $x$, irrespective of the type of interference signals. The architecture of AV-GridNet is illustrated in Fig.~\ref{fig:av-gridnet}, incorporating an additional visual conditioning network to extract visual features $V$ from the face recording $v$.

\vspace{-2mm}
\subsubsection{Visual conditioning network}
\vspace{-2mm}
The visual conditioning network comprises a 3-dimensional convolutional layer (Conv3D), a ResNet 18 layer, and $R$ repetitions of visual temporal convolutional network (V-TCN), as depicted in Fig.~\ref{fig:av-gridnet}. The Conv3D and ResNet 18 layers are pre-trained on lip-reading tasks and are kept frozen during the training of AV-GridNet\footnote{The pre-trained visual network can be found at \url{https://github.com/smeetrs/deep_avsr}~\cite{Afouras18b}}. This allows the network to retain the ability to encode viseme movements that synchronize with the phoneme sequence of speech. Additionally, we employ V-TCN layers as an adaptation, similar to~\cite{pan2021reentry,usev21}, to adapt the visual embeddings towards speech extraction. 

The visual embedding $V$ typically has a lower temporal resolution compared to speech embeddings. To address this mismatch, we linearly interpolate the visual embeddings along the time dimension to match the resolution of the speech embeddings. We fuse the same visual embeddings $V$ using a fusion layer to the start of each GridNet block. Specifically, we concatenate the audio and visual embeddings along the channel dimension, and project them back to the original channel dimension of the audio embeddings before fusion with a linear layer.

\vspace{-2mm}
\subsubsection{Loss function to train the AV-GridNet}
\vspace{-2mm}
In time-domain end-to-end speaker extraction network training, the negative SI-SDR loss function~\cite{le2019sdr} has been widely used in most methods. It is formulated as follows:
\begin{equation}
    \label{eqa:loss_sisnr}
    \mathcal{L}_{\text{SI-SDR}}(s, \hat{s}) = - 20 \log_{10} \frac{\big\|\frac{<\hat{s},s>}{\|s\|^2}s\big\|}{\big\|\hat{s} - \frac{<\hat{s},s>}{\|s\|^2}s\big\|}.
\end{equation}
In this work, we adopt the hybrid loss proposed in~\cite{pan2022hybrid}, which reduces the over-suppression error and leads to improved perceptual quality and intelligibility for the extracted speech. The hybrid loss consists of the time-domain SI-SDR loss as shown in Eq.~(\ref{eqa:loss_sisnr}), together with a frequency-domain multi-resolution delta spectrum loss\footnote{Code for the hybrid loss function can be found at~\url{https://github.com/zexupan/avse_hybrid_loss}~\cite{pan2022hybrid}}:
\begin{equation}
\mathcal{L}_{\text{hybrid}}(s, \hat{s})=\mathcal{L}_{\text{SI-SDR}}(s, \hat{s})+\gamma \frac{1}{M}\sum^M_{m=1}\mathcal{L}_{\text{freq-$\Delta$}}^m (s, \hat{s}),
\end{equation}
where the delta spectrum loss $\mathcal{L}^m_{\text{freq-$\Delta$}}$ is calculated at $M=3$ different resolutions, using the following triplets of parameters for \{FFT size, hop size, window length\} in samples: \{512, 50, 240\}, \{1024, 120, 600\}, and \{2048, 240, 1200\}. %
% with number of FFT bins $\in$ \{512, 1024, 2048\}, hop sizes $\in$ \{50, 120, 240\}, and window lengths $\in$ \{240, 600, 1200\} respectively. 
$\gamma$ is a balancing weight that is set to 1 in this paper.

\vspace{-2mm}
\subsection{Proposed scenario-aware SAV-GridNet}
\vspace{-2mm}
\subsubsection{Motivation}
\vspace{-2mm}
Speech and noise exhibit very distinct characteristics, and scenarios where one or the other acts as interfering signal of a target speaker may thus require different strategies.
Indeed, a model trained to separate speech from speech is likely to heavily rely on the structure of speech, and thus to be different from a model trained to separate speech from noise, which has the opportunity to rely on the intrinsic differences between the characteristics of the two signals to be separated.
Therefore, we advocate that a dedicated expert AV-GridNet model that is trained specifically for noise or speech interference may better handle each scenario.

\begin{figure}
  \centering
  \includegraphics[width=0.99\linewidth]{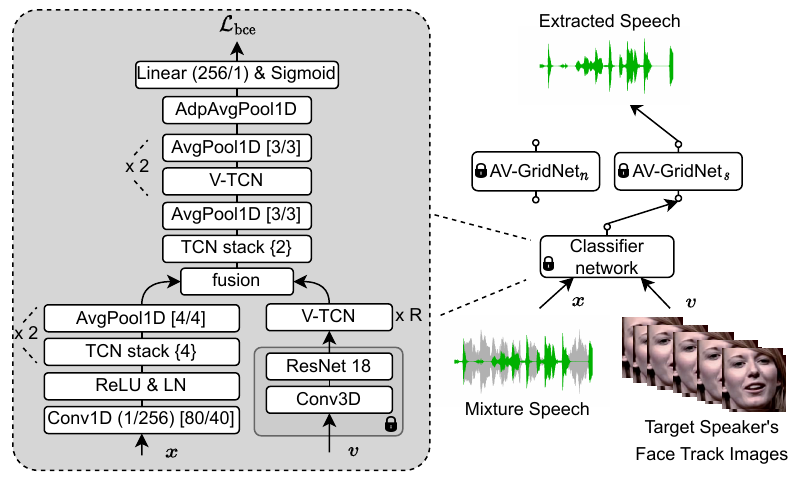}
  \vspace{-4mm}
  \caption{Our scenario-aware SAV-GridNet model is a cascaded model that first classifies the type of interfering signals with a classifier network, and then applies dedicated expert models AV-GridNet$_n$ for noise interference or AV-GridNet$_s$ for speech interference. 
  In the classifier network, the values inside ``(/)'' represent the input and output feature sizes, the values inside ``[/]'' represent kernel size and stride, and the value inside ``\{\}'' represents the upper-bound of the dilation value in a TCN stack~\cite{luo2019conv}. %ReLU denotes rectified linear unit activation,  %%% that should be obvious by now
  AvgPool1D and AdpAvgPool1D denote 1D average pooling and 1D adaptive average pooling operations, respectively.}
  \label{fig:sav-gridnet}
  \vspace{-3mm}
\end{figure}

To this end, we propose a model, referred to as SAV-GridNet, that is aware of the different interference scenarios as depicted in Fig.~\ref{fig:sav-gridnet}. SAV-GridNet first identifies the type of interfering scenario with a classifier network, and then applies a dedicated expert model that is trained specifically for that scenario. The classifier model and the expert models AV-GridNet$_n$ (for noise interference) and AV-GridNet$_s$ (for speech interference) are trained independently.

Note that we consider here scenarios involving either noise or a single speaker as interference because of the particular setting of the COG-MHEAR challenge. A generalized and arguably more realistic setting for practical applications would be to consider noise-only interference on one hand, and one or more speakers with or without background noise on the other. While our proposed classifier-based approach can be readily extended to this setting, with the corresponding expert models, we leave a thorough investigation of the performance of such a system to future work.

\vspace{-2mm}
\subsubsection{Classifier network}
\vspace{-2mm}
The classifier network, detailed in the left panel of Fig.~\ref{fig:sav-gridnet}, accepts both $x$ and $v$ as inputs. Although the task could be performed with only $x$, it may be beneficial to include the visual signals here as it could serve as an anchor point for the target speech. The classifier network design is motivated by the audio-visual SLSyn network in~\cite{pan2021reentry}, which consists of a visual front-end, a speech front-end, and an audio-visual back-end. It is worth noting that the visual front-end here also consists of Conv3D, ResNet 18, and V-TCN layers. As with the visual conditioning network in Fig.~\ref{fig:av-gridnet}, the Conv3D and ResNet 18 layers are pre-trained on lip-reading tasks and are kept frozen during the training of the classifier model.

We minimize the following binary cross-entropy loss for the scenario classifier network training:
\begin{equation}
    \mathcal{L}_{\text{bce}} = - y\log(\hat{y}) - (1-y)\log(1-\hat{y}),
\end{equation}
where $y \in \{0,1\}$ indicates whether the interfering signal is speech or noise, while $\hat{y}$ is the predicted probability. We arbitrarily set speech interference to be the negative class and noise interference to be the positive class.

\vspace{-2mm}
\subsubsection{Classifier post-processing}
\vspace{-2mm}

If the classifier makes a mistake and the wrong expert model is used, the results may be detrimental as there is a mismatch between training and inference for the AV-GridNet. We empirically find that a model trained only on speech interference generalizes remarkably well on noise interference, but not vice versa. Therefore, we propose two post-processing strategies to mitigate the false-positive cases (i.e., predicting noise while the ground-truth label is speech).

For the first post-processing strategy (post-proc$_1$), if the classifier prediction is noise and the criteria in Eq.~(\ref{eqa:pos_1}) is met, which indicates that the universal model is more in agreement with the noise expert than with the speech expert, we classify the interference as noise, otherwise as speech:
\begin{equation}
    \label{eqa:pos_1}
    \mathcal{L}_{\text{SI-SDR}}(\hat{s}, \hat{s}_n) < \mathcal{L}_{\text{SI-SDR}}(\hat{s}, \hat{s}_s),
\end{equation}
where
\begin{align}
    \hat{s}_n &= \texttt{AV-GridNet$_n$}(x, v ),\\
    \hat{s}_s &= \texttt{AV-GridNet$_s$}(x, v ),\\
    \hat{s} &= \texttt{AV-GridNet}(x, v ).
\end{align}

For the second post-processing strategy (post-proc$_2$), if the classifier prediction is noise, and either of the criteria in Eq.~(\ref{eqa:pos_1}) or Eq.~(\ref{eqa:pos_2}) is met, with this latter criterion indicating that the original mixture is further to the output of the noise expert than to that of the speech expert, we classify the interference as noise, otherwise as speech:
\begin{equation}
    \label{eqa:pos_2}
    \mathcal{L}_{\text{SI-SDR}}(x, \hat{s}_n) > \mathcal{L}_{\text{SI-SDR}}(x, \hat{s}_s)
\end{equation}

For the samples that are classified as noise interference initially by the classifier but are changed to speech interference by the post-processing, we use the extracted speech from the universal AV-GridNet model. For all other samples, the model indicated by the classifier is used.

\vspace{-2mm}
\section{EXPERIMENTAL SETUP}
\vspace{-2mm}
\subsection{Dataset}
\vspace{-2mm}
We participated in the second COG-MHEAR Audio-Visual Speech Enhancement Challenge~\footnote{\url{https://challenge.cogmhear.org/}} and evaluated our proposed method on its benchmark dataset. The speech dataset is from the Lip Reading Sentences 3 (LRS3)~\cite{afouras2018lrs3}, which consists of thousands of spoken sentences from TED videos. The noise datasets are from the Clarity challenge~\cite{graetzer2021clarity}, which comprises around $7$ hours of domestic noises, the DEMAND dataset~\cite{thiemann2013demand}, which includes recordings of 18 soundscapes that represent over $1$ hour of data, and the Deep Noise Suppression (DNS) challenge~\cite{dubey2023icassp}, for which only the noise signals from Freesound~\cite{fonseca2021fsd50k} are considered. The challenge has two tracks: systems in track $1$ can only use the above-mentioned provided datasets and unimodal pre-trained models, while systems in track $2$ have no limitations in the datasets and pre-trained models used.

The training, development, and evaluation sets consist of $34519$, $3300$, and $2792$ scenes respectively. There are two scenarios in total, a target speaker mixed with a competing speaker at random signal-to-noise ratio (SNR) levels that range from $-15$ dB to $+5$ dB, or a target speaker mixed with a noise signal at random SNR levels that range from $-10$ dB to $+10$ dB. The clean speech signals and scenario labels are only available for the training and development set. The audio signals are sampled at $16$~kHz, while the video has a frame rate of $25$ per second. The target face tracks are provided for all the samples.

\vspace{-2mm}
\subsection{Baselines}
\vspace{-2mm}

We use the AV-DPRNN network~\cite{pan2022hybrid,usev21} as our main baseline, as it is currently one of the best-performing audio-visual speech extraction networks. There are three main differences between AV-DPRNN and AV-GridNet: 1) The speech encoder and decoders used by AV-DPRNN are in the time domain, while those of AV-GridNet are in the time-frequency domain; 2) AV-DPRNN is a mask-based method that uses dual-path BLSTM as the extractor, while AV-GridNet directly maps the signals using the GridNet blocks as the extractor; 3) The visual embeddings are only fused at the first repeat of the extractor for AV-DPRNN, while the visual embeddings are fused at every repeat of the extractor for AV-GridNet.
We also report results by the official baseline released by the challenge organization. It has a similar architecture to the AV-DPRNN network~\cite{pan2022hybrid,usev21}, but with no visual pre-training involved.

\vspace{-2mm}
\subsection{Model and training settings}
\vspace{-2mm}

For the baseline AV-DPRNN, the hyperparameter setting follows~\cite{pan2022hybrid,usev21}. For the classifier network in SAV-GridNet, the TCN stack hyperparameter follows~\cite {luo2019conv}. For AV-GridNet, the V-TCN hyperparameter follows~\cite{pan2022hybrid,usev21}. We set $D=48$, $B=6$, and $R=5$. The STFT window size is $256$, the hop size is $128$, and the square root Hann window is used. A $256$-point discrete Fourier transform is applied to extract $129$-dimensional complex spectra at each frame. For other hyperparameters in the GridNet block, we set $I=4$, $J=1$, $H=192$, $E=4$, and $L=4$~\cite{wang2023tf}.

For all model training, we use the Adam optimizer with an initial learning rate of 0.001, the learning rate is halved if the best development loss (BDL) does not improve for 6 consecutive epochs, and the training stops when the BDL does not improve for 20 consecutive epochs. We train the model on $8$ GPUs with \SI{48}{\giga\byte} RAM each. To fit the data in the GPU memory during training, the audio clips are truncated to 3 seconds for AV-GridNet, 12 seconds for AV-DPRNN, and 25 seconds for the classifier network.

\begin{table*}
    \centering
    \sisetup{
      mode=text, % Make siuntix print tables in text mode (causes width of bold characters to be the same as non-bold)
      table-format=1.2,
      round-mode=places,
      round-precision=2,
      table-number-alignment = center,
      detect-weight=true,
      tight-spacing=true}
    \caption{Development set results on the 2nd COG-MHEAR Audio-Visual Speech Enhancement Challenge benchmark. Experiments are done for track $1$, in which only the provided dataset is used in training. We use the system number (Sys.) to identify different systems. Init. indicates which (if any) other system is used to initialize that system's parameters (note that the optimizer parameters are always re-initialized). DM stands for dynamic mixing, which involves simulating the mixture utterances on the fly during training using the protocol provided by the challenge organizers. We report performance on the~speech+speech scenario, the speech+noise scenario, and overall.
    % Note that SAV-GridNet* (sys. 10) used the oracle scenario labels to demonstrate the performance upper bound of our SAV-GridNet.
    The symbol * indicates the model used the oracle scenario labels.
    }
    % \addtolength{\tabcolsep}{-3.5pt}
    \vspace*{.5mm}
    \resizebox{0.99\linewidth}{!}
    {
    \begin{tabular}{c*{4}{c}*{3}{SSS[round-precision=1,table-format=2.1]}} 
       \toprule
        % \multirow{2}*{Sys.}    &\multirow{2}*{Model}   &\multirow{2}*{Init.}       &\multirow{2}*{DM}    &\multirow{2}*{Loss} 
        &&&&
        &\multicolumn{3}{c}{Speech+Speech} &\multicolumn{3}{c}{Speech+Noise}  &\multicolumn{3}{c}{Overall}\\
        \cmidrule(lr){6-8} \cmidrule(lr){9-11} \cmidrule(lr){12-14}
        %&&&&
        Sys.&Model&Init.&DM&Loss
        &{PESQ}   &{STOI}   &{SI-SDR}   &{PESQ}   &{STOI}   &{SI-SDR} &{PESQ}   &{STOI}   &{SI-SDR}   \\
        \midrule
        -&Noisy &-  &- &-
                &1.166	&0.596	&-5.02	  &1.147  &0.6783 &-4.37  &1.156  &0.638  &-4.69\\
        \midrule
        1& \multirow{2}*{AV-DPRNN}  &- &\multirow{2}*{\xmark}  &$\mathcal{L}_{\text{SI-SDR}}$
                &1.677	&0.839	&8.84     &1.710  &0.823  &9.83   &1.694  &0.831  &9.34\\
        2&  &1 & &   $\mathcal{L}_{\text{hybrid}}$
                &2.227  &0.896  &12.57    &2.022    &0.860  &11.44  &2.124	&0.877	&12.00\\
        \midrule
        3&\multirow{2}*{AV-GridNet} &- &\xmark &\multirow{2}*{$\mathcal{L}_{\text{hybrid}}$}
                &3.066	&0.945	&17.27		&2.355	&0.869	&12.84   &2.777	&0.922	&14.61\\
        4&  &3 &\cmark   &
                &3.097	&0.946	&16.68		&2.619	&0.907	&13.92   &2.857	&0.927	&15.29\\
        \midrule
        5&AV-GridNet$_s$    &\multirow{2}*{3}  &\multirow{2}*{\cmark}&\multirow{2}*{$\mathcal{L}_{\text{hybrid}}$}
                &3.229	&0.952	&17.51		&2.557	&0.900	&13.44 &2.891	&0.926	&15.46\\
        6&AV-GridNet$_n$    &   &  &
                &1.266	&0.610	&-4.69		&2.681	&0.913	&14.18 &1.979	&0.762	&4.81\\
        \midrule
        7&SAV-GridNet   &\multirow{3}*{-}  &\multirow{3}*{\cmark}&\multirow{3}*{$\mathcal{L}_{\text{hybrid}}$}
                &3.220	&0.950	&17.39		&2.681	&0.913	&14.18 &2.949	&0.931	&15.78\\
        8& + post-proc$_1$ & &&      
                &3.227	&0.952	&17.50		&2.677	&0.912	&14.16 &2.951	&0.932	&15.82\\
        9& + post-proc$_2$ & &&        
                &3.228	&0.952	&17.50		&2.679	&0.912	&14.16 &2.951	&0.932	&15.82\\
        \midrule
        10&SAV-GridNet* &-  &\cmark &$\mathcal{L}_{\text{hybrid}}$
                &3.229	&0.952	&17.51		&2.681	&0.913	&14.18 &2.953	&0.932	&15.84\\
        \bottomrule
    \end{tabular}
    }
    % \addtolength{\tabcolsep}{3.5pt}
    \vspace*{-3mm}
    \label{tab:dev}
\end{table*}

\vspace{-2mm}
\section{RESULTS}
\vspace{-2mm}
We evaluate the speech signals extracted by our proposed networks and the baselines using objective measures  PESQ, STOI, and SI-SDR. PESQ measures the perceptual quality of the extracted speech signal and is in the range of $-0.5$ to $4.5$; STOI measures the intelligibility of the extracted speech signal and is in the range of $0$ to $1$; and SI-SDR measures the signal quality of the extracted speech signal in dB and is unbounded. The higher the better for all three metrics. We use PESQ as our main measure when describing the results, as other measures show similar trends.

\vspace{-2mm}
\subsection{Comparison with baseline and ablation study}
\vspace{-2mm}

In Table~\ref{tab:dev}, we present the results of our baseline and proposed models on the development set. For our baseline AV-DPRNN, it is seen that using hybrid loss (Sys.\ 2) outperforms (Sys.\ 1) by 0.43 on the overall PESQ, showing the effectiveness of the frequency-domain loss on the speech perceptual quality.

Our first proposed AV-GridNet (Sys.\ 3) outperforms AV-DPRNN (Sys.\ 2) by $0.66$ for PESQ. With additional dynamic mixing (Sys.\ 4), the PESQ further improves by $0.08$.
Our expert models, AV-GridNet$_s$ and AV-GridNet$_n$, improved PESQ compared to AV-GridNet by $0.13$ in the speech+speech scenario and by $0.06$ in the speech+noise scenario, respectively.
Thanks to adaptively choosing the expert models, our proposed SAV-GridNet (Sys.\ 7) outperforms the AV-GridNet (Sys.\ 4) regardless of the scenarios.
While the post-processing techniques (Sys.\ 8 and 9) do not appear to show performance improvements on the averaged metrics, they do reduce outliers, as will be investigated in the next subsection.
The performance of our best system is nearly identical to that of SAV-GridNet with the oracle scenario labels (Sys.\ 10).

\begin{figure}[t]
    \begin{minipage}[t]{.32\linewidth}
      \centering
      \centerline{\includegraphics[width=\linewidth]{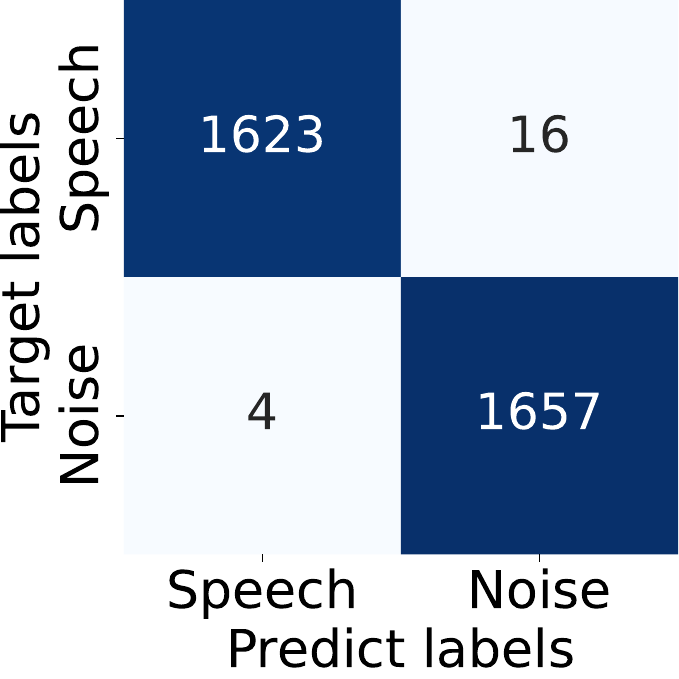}}
      \centerline{\scalebox{0.8}{(a) System 7}}\medskip
      \vspace*{-4mm}
    \end{minipage}
\hfill
    \begin{minipage}[t]{.32\linewidth}
      \centering
      \centerline{\includegraphics[width=\linewidth]{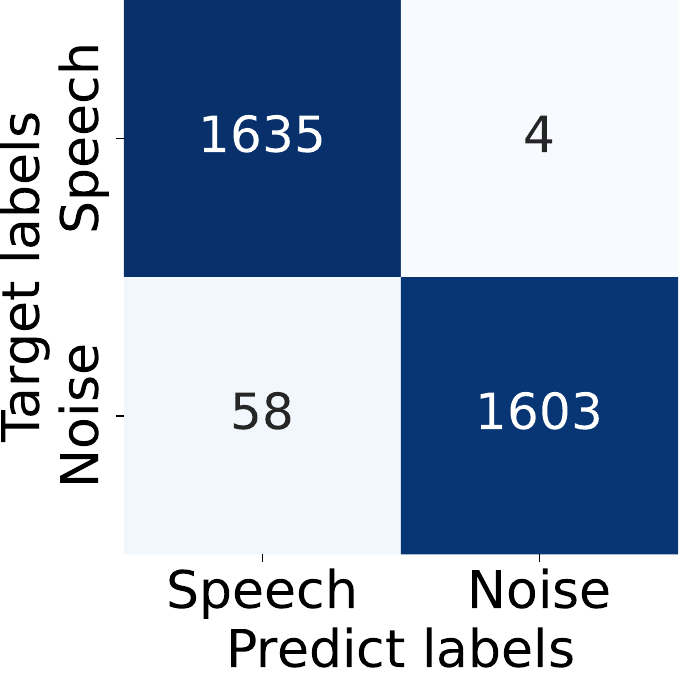}}
      \centerline{\scalebox{0.8}{(b) System 8}}\medskip
      \vspace*{-4mm}
    \end{minipage}
\hfill
    \begin{minipage}[t]{.32\linewidth}
      \centering
      \centerline{\includegraphics[width=\linewidth]{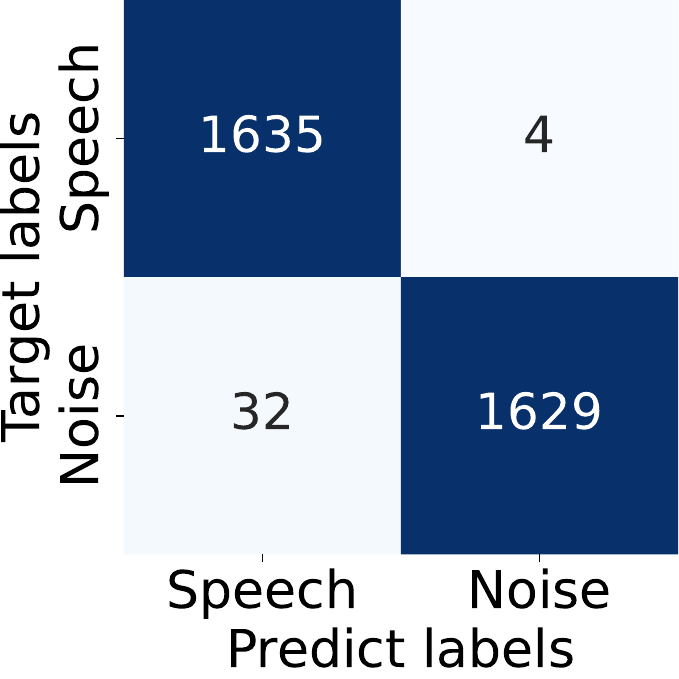}}
      \centerline{\scalebox{0.8}{(c) System 9}}\medskip
      \vspace*{-4mm}
    \end{minipage}
\caption{Confusion matrix of the scenario classification on the development set.}
\vspace*{-4mm}
\label{figure:cm}
\end{figure}

\vspace{-2mm}
\subsection{Analysis for speech and noise interfering signals}
\vspace{-2mm}

Fig.~\ref{figure:cm} illustrates the confusion matrix of the scenario classification with and without post-processing.
Our scenario classification network achieved accuracy over $99\%$ without post-processing.
Since the false-positive cases (predicting noise when the ground-truth label is speech) severely deteriorate the subsequent target speech extraction performance according to Sys.\ 6 in Table~\ref{tab:dev}, it is important to reduce the number of false-positive cases, which post-proc$_1$ does successfully.
This however increased the false-negative cases to $58$, but post-proc$_2$ mitigated this increase and performed best overall in terms of target speech extraction.

\begin{figure}[t]
\begin{minipage}[t]{.49\linewidth}
  \centering
  \centerline{\includegraphics[width=\linewidth]{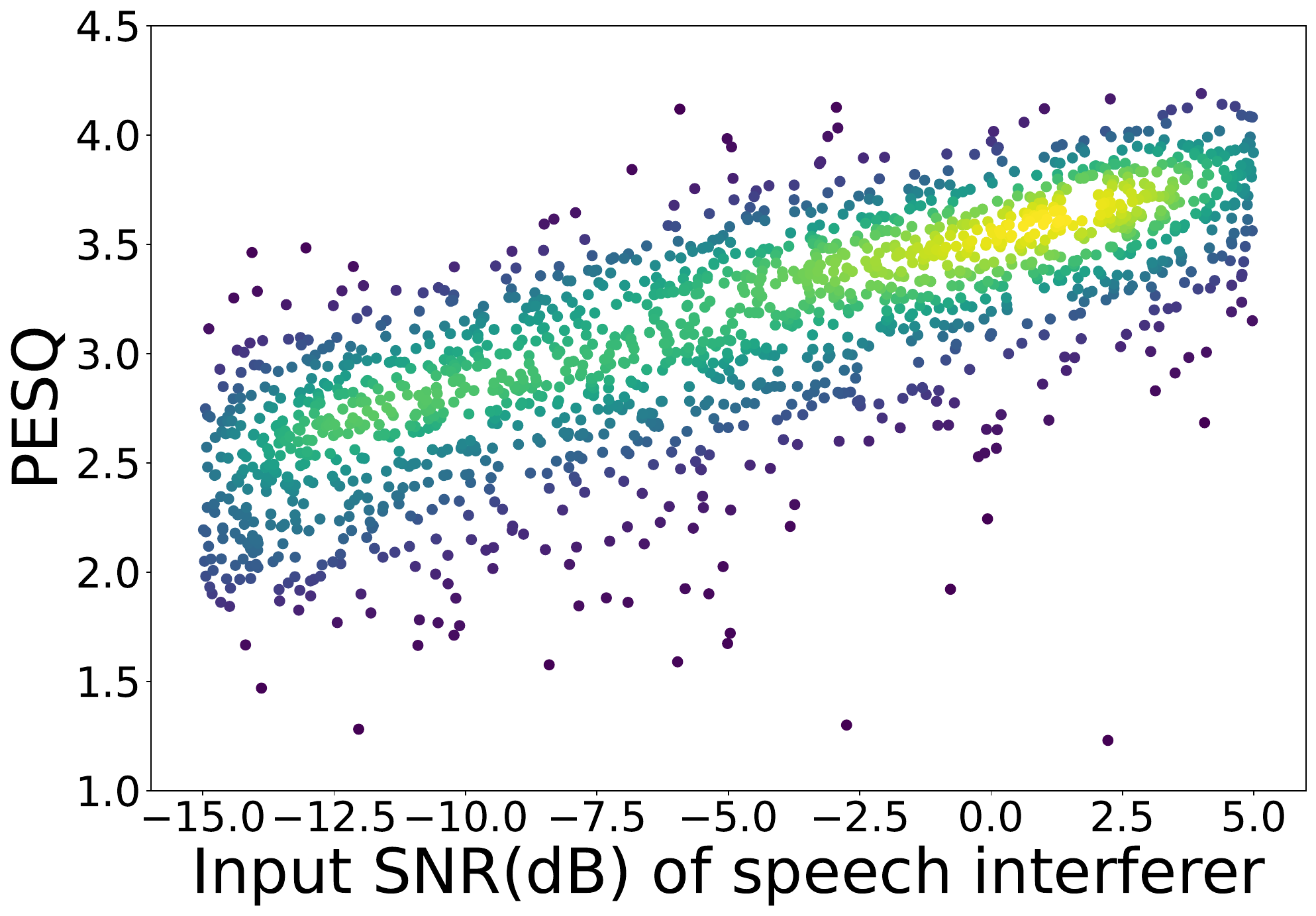}}
  % \vspace*{-2mm}
\end{minipage}
\hfill
\begin{minipage}[t]{.49\linewidth}
  \centering
  \centerline{\includegraphics[width=\linewidth]{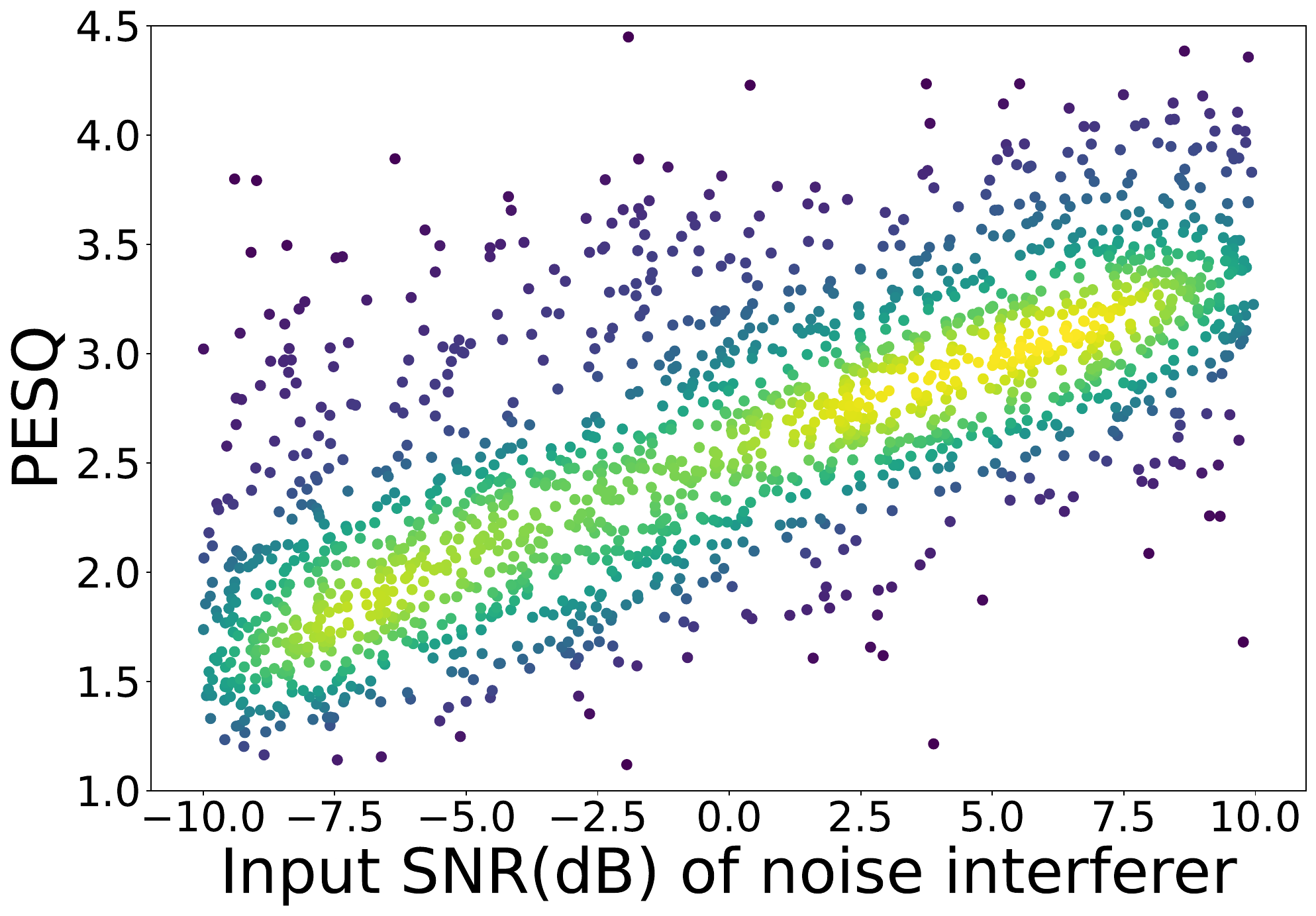}}
  % \vspace*{-2mm}
\end{minipage}
\vspace{-2mm}
\caption{PESQ of extracted speech signal from system 4 for the speech (left) and noise (right) interfering signals.}
\vspace{-2mm}
\label{figure:pesq_sys4}
\end{figure}

\begin{figure}[t]
\begin{minipage}[t]{.49\linewidth}
  \centering
  \centerline{\includegraphics[width=\linewidth]{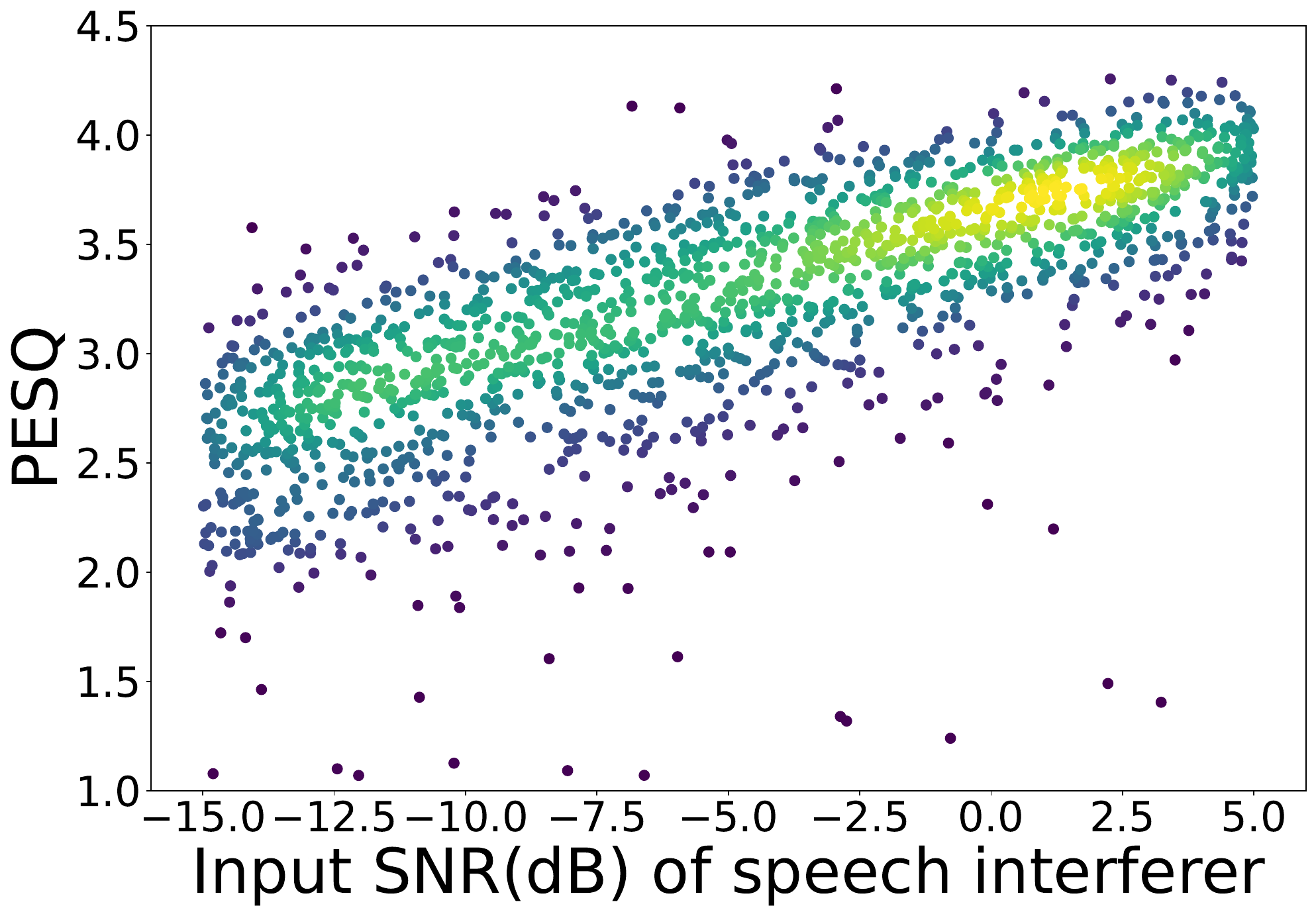}}
  % \vspace*{-2mm}
\end{minipage}
\hfill
\begin{minipage}[t]{.49\linewidth}
  \centering
  \centerline{\includegraphics[width=\linewidth]{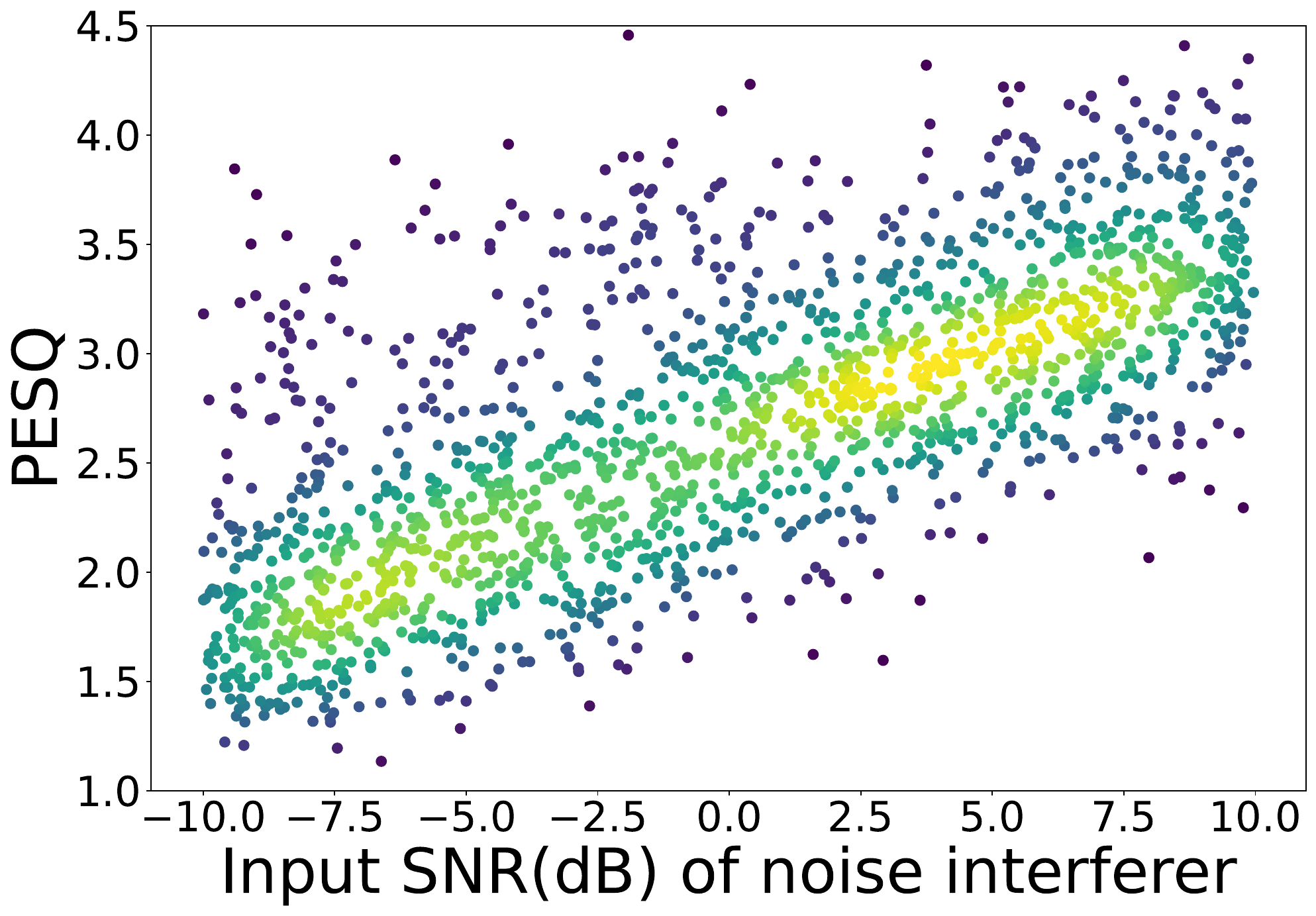}}
  % \vspace*{-2mm}
\end{minipage}
\vspace{-2mm}
\caption{PESQ of extracted speech signal from system 7 for the speech (left) and noise (right) interfering signals.}
\vspace{-2mm}
\label{figure:pesq_sys7}
\end{figure}

\begin{figure}[t]
\begin{minipage}[t]{.49\linewidth}
  \centering
  \centerline{\includegraphics[width=\linewidth]{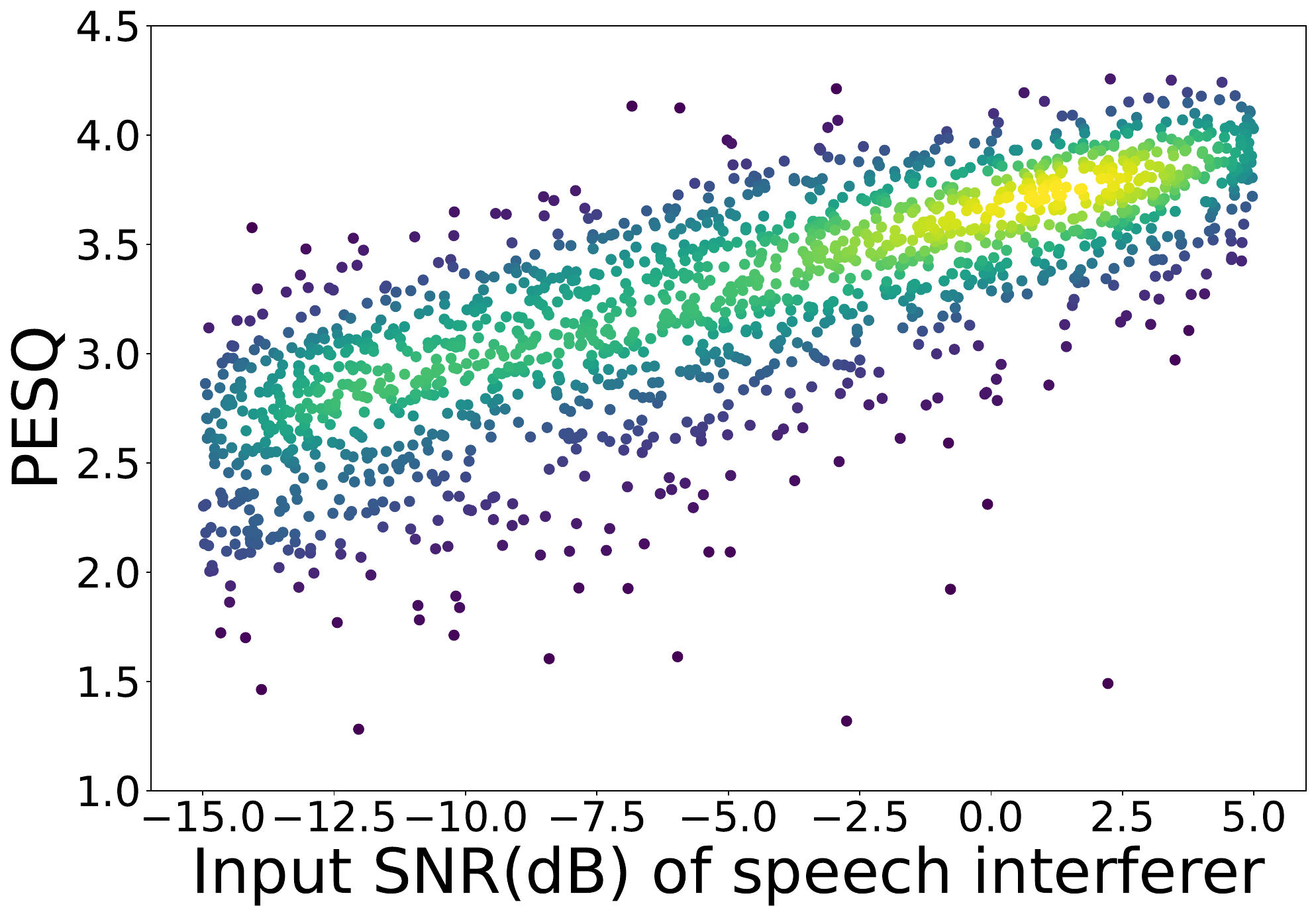}}
  % \vspace*{-2mm}
\end{minipage}
\hfill
\begin{minipage}[t]{.49\linewidth}
  \centering
  \centerline{\includegraphics[width=\linewidth]{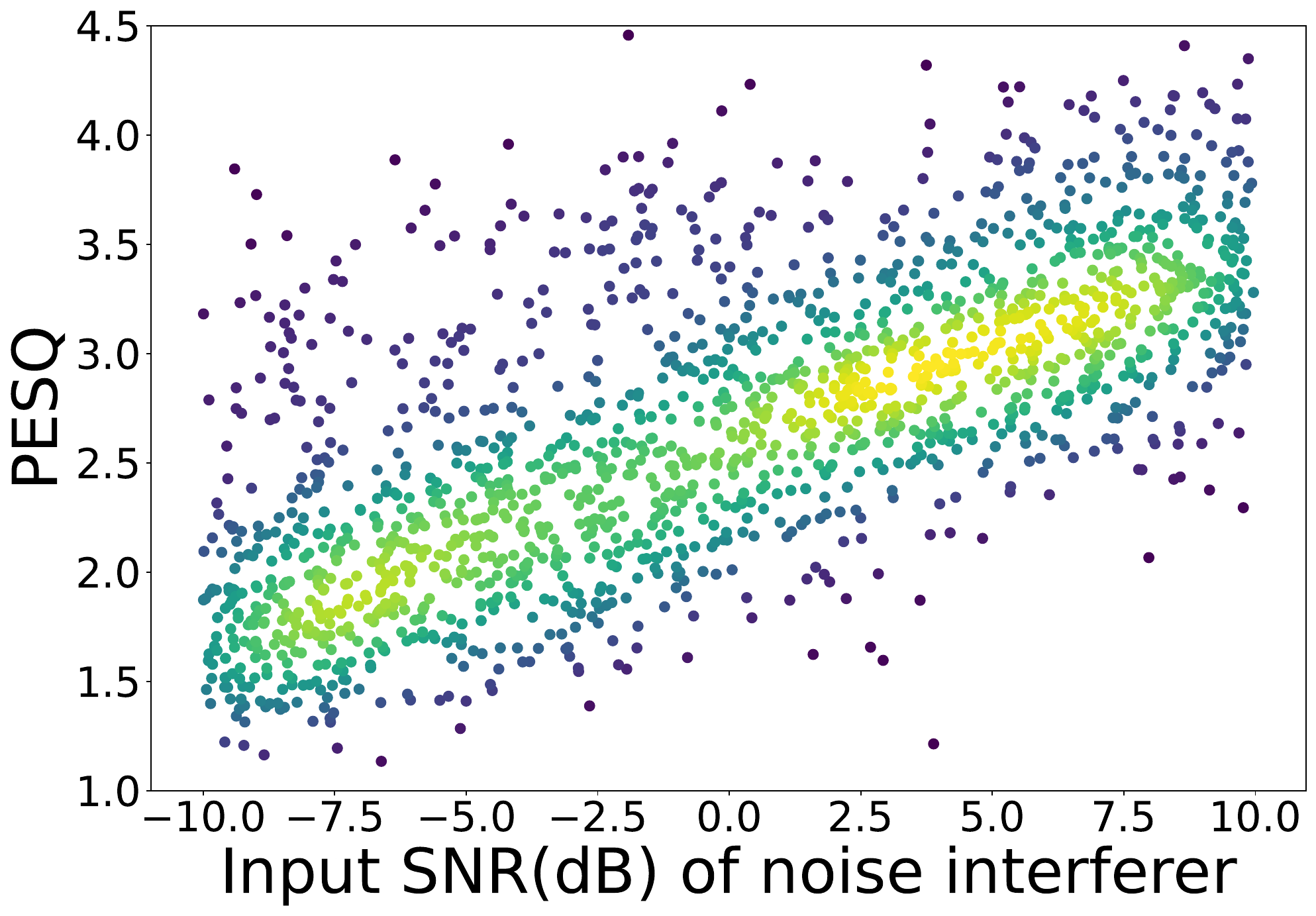}}
  % \vspace*{-2mm}
\end{minipage}
\vspace{-2mm}
\caption{PESQ of extracted speech signal from system 9 for the speech (left) and noise (right) interfering signals.}
\label{figure:pesq_sys9}
\vspace{-2mm}
\end{figure}

This tendency is also confirmed from the distributions of PESQ shown in Figs.~\ref{figure:pesq_sys4}--\ref{figure:pesq_sys9}.
SAV-GridNet (Sys.\ 7, Fig.~\ref{figure:pesq_sys7}) improved the overall performance from AV-GridNet (Sys.\ 4, Fig.~\ref{figure:pesq_sys4}), but it had more outliers in the speech+speech scenario.
This is likely because AV-GridNet$_n$ was applied to some speech+speech samples due to misclassification.
As the post-processing techniques successfully reduced the false-positive cases, Sys.\ 9 reduced the number of outliers with low PESQ in the left panel of Fig.~\ref{figure:pesq_sys9}, while substantially preserving the distribution in the speech+noise scenario. Overall, the number of samples with a PESQ value smaller than $1.5$ went down from $62$ for Sys.\ 7 to $54$ and $53$ for Sys.\ 8 and 9, respectively, post-proc$_1$ thus reducing the number of such failing samples by 13\%, and post-proc$_2$ by 14.5\%.

\begin{figure}[t]
\begin{minipage}[t]{.48\linewidth}
  \centering
  \centerline{\includegraphics[width=\linewidth]{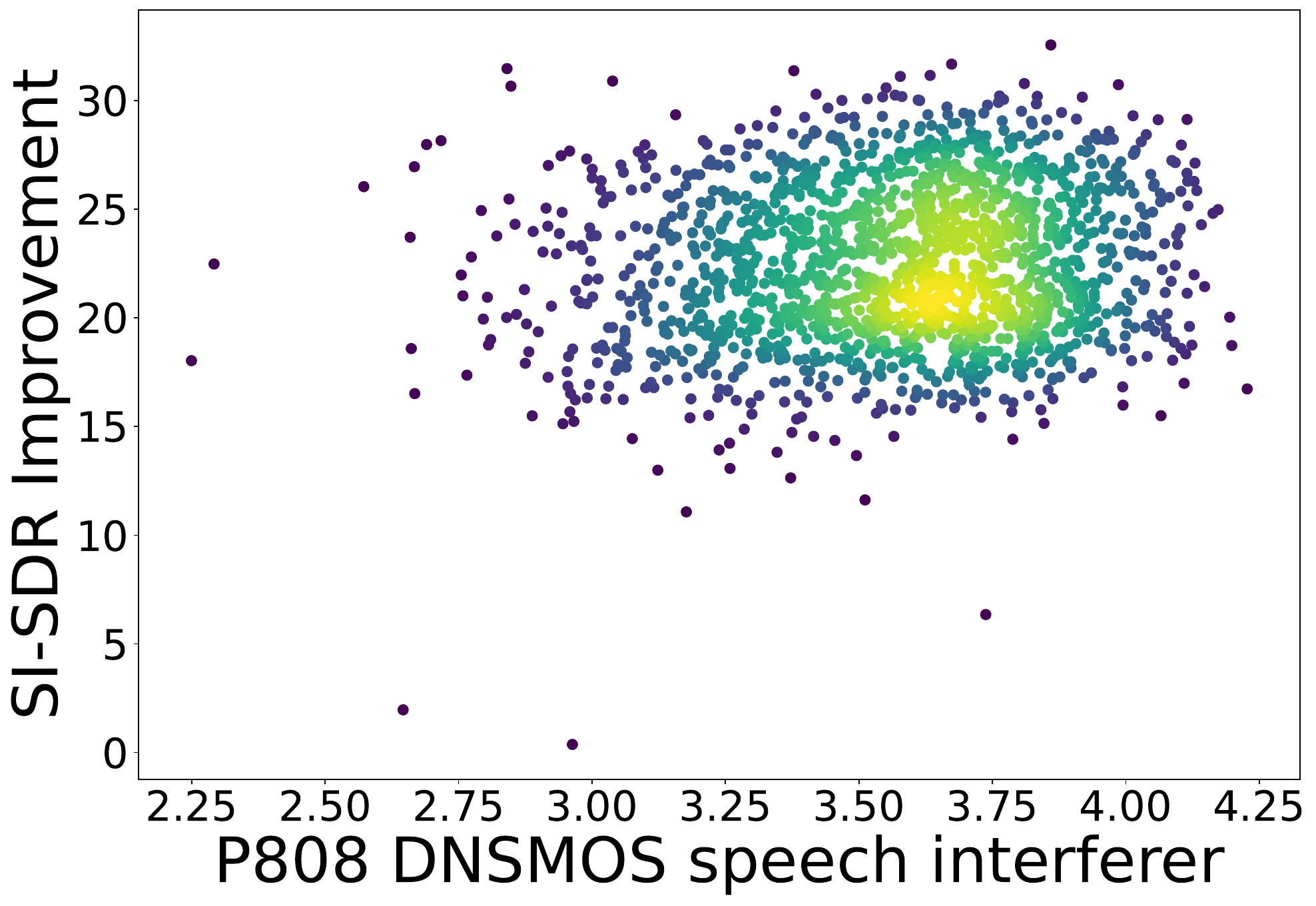}}
  % \vspace*{-2mm}
\end{minipage}
\hfill
\begin{minipage}[t]{.49\linewidth}
  \centering
  \centerline{\includegraphics[width=\linewidth]{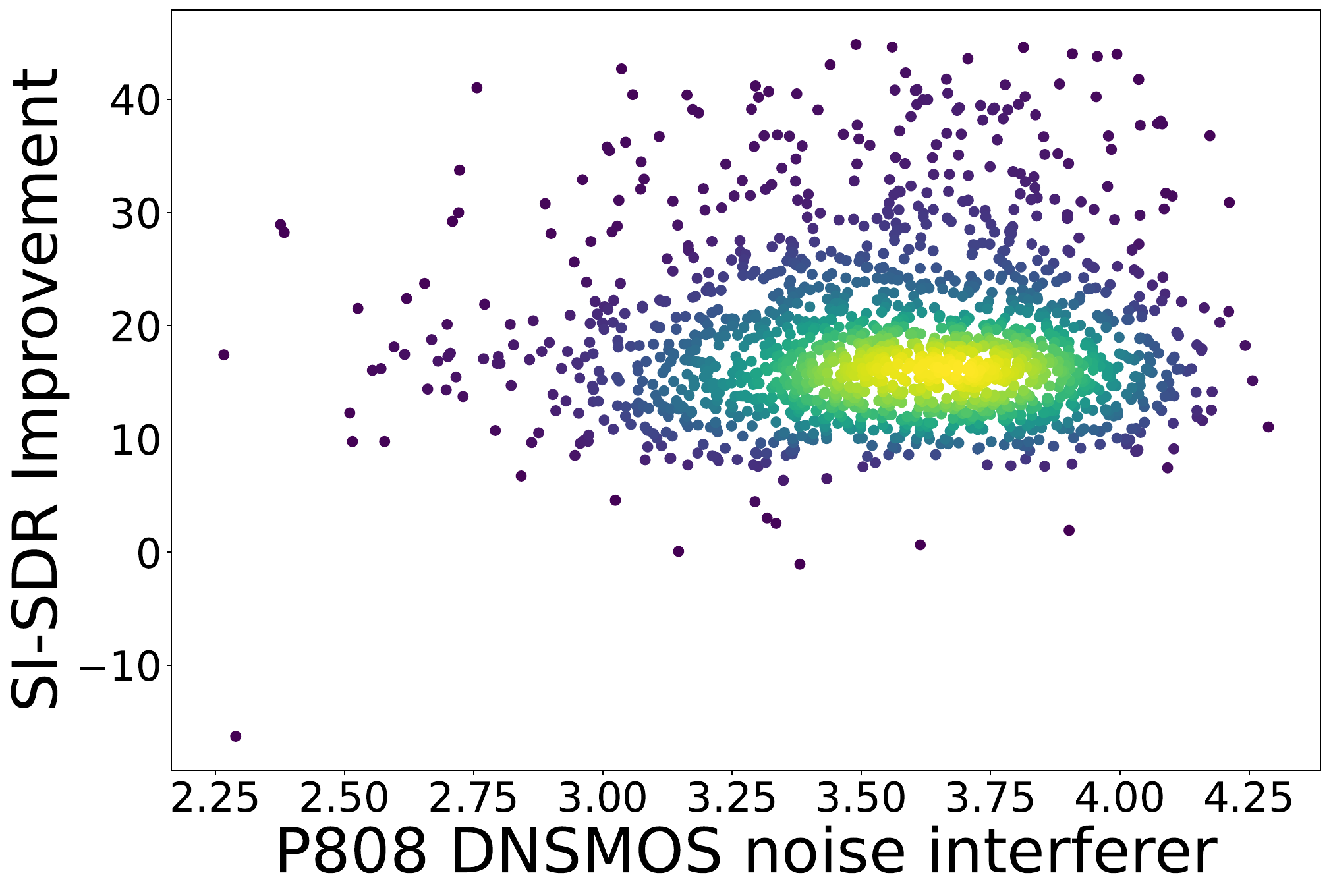}}
  % \vspace*{-2mm}
\end{minipage}
\vspace{-2mm}
\caption{Development set SI-SDR improvement [dB] vs.\ P808 DNSMOS for system 9 for speech (left) and noise (right) interfering signals. Many of the lowest performing samples appear to have noisy ground-truth signals based on their low score under the reference-free DNSMOS quality measure.}
\label{figure:dnsmos_outliers}
\vspace{-2mm}
\end{figure}

\noindent{\bf Analyzing low performing samples}: When informally listening to samples with the lowest objective metrics, we noticed that, while a few samples did have mid-utterance switching between target and interfering speakers in the speech+speech scenario, the main issue was that many of the target speech signals were not very clean, e.g., they contained impulsive disturbances from microphone contact or crowd noises such as cheering and clapping. To quantify objectively the quality of the target speech signals, we used the P808 DNSMOS~\cite{reddy2021dnsmos} score, a reference-free measure for evaluating overall audio quality. Figure~\ref{figure:dnsmos_outliers} displays the distribution of SI-SDR vs. DNSMOS. We chose SI-SDR over PESQ as the reference metric for this figure because the larger dynamic range makes outliers more visible. 
We note that all of the lowest SI-SDR samples output by our model have DNSMOS values below $3.0$ in Fig.~\ref{figure:dnsmos_outliers}, and we found by informal listening that these samples contained noisy target speech. As datasets become larger, using a reference-free speech quality metric could help remove noisy target speech samples.

\vspace{-2mm}
\subsection{Performance on leaderboard}
\vspace{-2mm}

In Table~\ref{tab:evl}, we present the performance of our models on the hidden evaluation set, for which the numbers are obtained from the submissions to the leaderboard\footnote{\url{https://challenge.cogmhear.org/\#/results}}.
%\footnote{As of July 17, 2023, 10pm AoE}. 
We can see that our AV-DPRNN baseline (Sys.\ 2) outperforms the challenge baseline by $0.53$ in terms of PESQ, thanks to the combined use of the presented pre-trained models, hybrid loss, and training settings. We also report the results of the top 3 other teams on track 1 in terms of PESQ. Results for track 2 are not reported as they did not improve upon those of track 1.

AV-GridNet outperforms AV-DPRNN by $0.65$ in terms of PESQ, while SAV-GridNet further outperforms AV-GridNet by $0.12$. Similarly to the results obtained on the development set, we cannot see a difference on the averaged metrics from using post-processing with SAV-GridNet (Sys.\ 8 and 9), but we hope that the number of outliers will again be reduced. This will need to be confirmed when/if the evaluation set is released.
Furthermore, in a listening test, we achieved an overall word intelligibility score of 84.54\%, compared to 57.56\% for the baseline and 80.41\% for the next best team. The Fisher's least significant difference (LSD) was 2.14\%, indicating that our model offered statistically significant intelligibility improvements compared to all other systems.

\begin{table}[t]
    \centering
     \sisetup{
      detect-weight=true,
      mode=text, % Make siuntix print tables in text mode (causes width of bold characters to be the same as non-bold)
      table-format=1.2,
      round-mode=places,
      round-precision=2,
      table-number-alignment = center,
      tight-spacing=true}
    \vspace{-2mm}
    \caption{Evaluation set results on the second COG-MHEAR Audio-Visual Speech Enhancement Challenge benchmark. We only present the top-3 teams on track 1 out of 8 other teams based on the PESQ ranking.}
    % \vspace*{-3mm}
    % \addtolength{\tabcolsep}{-3.5pt}
    \resizebox{0.9\linewidth}{!}{
    \begin{tabular}{*{3}{c}SSS[round-precision=1,table-format=2.1]} 
       \toprule
        Sys.    &Model  &Track    &{PESQ}   &{STOI}   &{SI-SDR}    \\
        \midrule
        \multirow{2}*{-}
        &Noisy       &\multirow{2}*{-}
                &1.136  &0.441  &-5.07\\
        &Baseline	&
                &1.414  &0.556  &3.67\\
        \midrule
        \multirow{3}*{-}
        &Team 1 &\multirow{3}*{1} %AVSE02
                &1.612240	&0.682139	&8.811185 \\
        &Team 2 &   %ict_avsu
                &1.658780	&0.676747	&6.726131 \\
        &Team 3 &   %\try6
                &1.764875	&0.712322	&7.683856 \\
        % &\BioASPCITI &
        %         & 1.407641 & 0.537798 & 3.610343 \\
        % &\ictavsu&
        %         &  1.654811 & 0.675288 & 6.706725 \\
        % &\rezzsl&
        %         & 1.278807 & 0.502772 & 2.620131 \\
        % &\ENUJHU &
        %         & 1.199859 & 0.465583 & 1.657748 \\
        % &\teamoh&
        %         & 1.277711 & 0.076207 & -51.514072 \\
        % &\teamtry&   
        %         & 1.764875 & 0.712322 & 7.683856 \\
        % &\teamwvl&   
        %         & \\
        % \midrule
        % \multirow{4}*{-}
        % &\teamtry   &\multirow{4}*{2}
        %         & 1.600692 & 0.676567 & 5.336684 \\
        % &\TTIC   &
        %         & 1.147212 & 0.512228 & -34.170490 \\
        % &\diffusion &
        %         & 1.446989 & 0.540302 & 2.783300 \\
        % &\rezzsl&
        %         & 1.244972 & 0.473104 & -29.378798 \\
        \midrule
        2&AV-DPRNN &\multirow{5}*{1}
                &1.937  &0.730  &10.35\\
        4&AV-GridNet        &
                &2.585  &0.827  &13.77\\
        7&SAV-GridNet       &
                &2.707  &0.841  &14.50\\
        8&+ post-proc$_1$   &
                &2.705  &0.839  &14.48\\
        9&+ post-proc$_2$   &
                &2.705  &0.840  &14.48\\
        \bottomrule
    \end{tabular}
    }
    % \addtolength{\tabcolsep}{3.5pt}
    % \vspace*{-6mm}
    \label{tab:evl}
\end{table}

\vspace{-2mm}
\section{Conclusion}
\vspace{-2mm}
In this work, we explored visually-grounded target speaker extraction based on the TF-GridNet separation architecture. Considering the different characteristics of noise and speech as interfering signals raised by the 2nd COG-MHEAR Audio-Visual Speech Enhancement Challenge, we proposed a scenario-aware model named SAV-GridNet that is capable to apply an expert model to individual scenarios independently. Experimental results show that the scenario-aware model generally improves the quality of the extracted speech, while reducing the number of samples with very low quality.

% \clearpage
\balance
\bibliographystyle{IEEEbib}
\bibliography{strings,refs}

\begin{thebibliography}{10}

\bibitem{pan2020multi}
Z.~Pan, Z.~Luo, J.~Yang, and H.~Li,
\newblock ``Multi-modal attention for speech emotion recognition,''
\newblock in {\em Proc. Interspeech}, 2020.

\bibitem{snyder2018x}
D.~Snyder, D.~Garcia-Romero, G.~Sell, D.~Povey, and S.~Khudanpur,
\newblock ``X-vectors: Robust {DNN} embeddings for speaker recognition,''
\newblock in {\em Proc. ICASSP}, 2018.

\bibitem{qian2021multi}
X.~Qian, M.~Madhavi, Z.~Pan, J.~Wang, and H.~Li,
\newblock ``Multi-target {DoA} estimation with an audio-visual fusion mechanism,''
\newblock in {\em Proc. ICASSP}, 2021.

\bibitem{wang2022predict}
J.~Wang, X.~Qian, and H.~Li,
\newblock ``Predict-and-{U}pdate network: Audio-visual speech recognition inspired by human speech perception,''
\newblock {\em arXiv preprint arXiv:2209.01768}, 2022.

\bibitem{tao2021someone}
R.~Tao, Z.~Pan, R.~K. Das, X.~Qian, M.~Z. Shou, and H.~Li,
\newblock ``Is someone speaking? {E}xploring long-term temporal features for audio-visual active speaker detection,''
\newblock in {\em Proc. ACM Multimedia}, 2021.

\bibitem{bronkhorst2000cocktail}
A.~W. Bronkhorst,
\newblock ``The cocktail party phenomenon: A review of research on speech intelligibility in multiple-talker conditions,''
\newblock {\em Acta Acust. United Acust.}, vol. 86, no. 1, pp. 117--128, 2000.

\bibitem{zmolikova2023neural}
K.~Zmolikova, M.~Delcroix, T.~Ochiai, K.~Kinoshita, J.~{\v{C}}ernock{\`y}, and D.~Yu,
\newblock ``Neural target speech extraction: An overview,''
\newblock {\em IEEE Signal Process. Mag.}, vol. 40, no. 3, pp. 8--29, 2023.

\bibitem{Chenglin2020spex}
C.~Xu, W.~Rao, E.~S. Chng, and H.~Li,
\newblock ``Sp{E}x: Multi-scale time domain speaker extraction network,''
\newblock {\em IEEE/ACM Trans. Audio, Speech, Lang. Process.}, vol. 28, pp. 1370--1384, 2020.

\bibitem{wang2019voicefilter}
Q.~Wang, H.~Muckenhirn, K.~Wilson, P.~Sridhar, Z.~Wu, J.~R. Hershey, R.~A. Saurous, R.~J. Weiss, Y.~Jia, and I.~L. Moreno,
\newblock ``{VoiceFilter}: Targeted voice separation by speaker-conditioned spectrogram masking,''
\newblock in {\em Proc. Interspeech}, 2019.

\bibitem{spex_plus2020}
M.~Ge, C.~Xu, L.~Wang, E.~S. Chng, J.~Dang, and H.~Li,
\newblock ``{SpEx+}: A complete time domain speaker extraction network,''
\newblock in {\em Proc. Interspeech}, 2020.

\bibitem{he2020speakerfilter}
S.~He, H.~Li, and X.~Zhang,
\newblock ``Speakerfilter: Deep learning-based target speaker extraction using anchor speech,''
\newblock in {\em Proc. ICASSP}, 2020.

\bibitem{xiao2019single}
X.~Xiao, Z.~Chen, T.~Yoshioka, H.~Erdogan, C.~Liu, D.~Dimitriadis, J.~Droppo, and Y.~Gong,
\newblock ``Single-channel speech extraction using speaker inventory and attention network,''
\newblock in {\em Proc. ICASSP}, 2019.

\bibitem{shi2020speaker}
J.~Shi, J.~Xu, Y.~Fujita, S.~Watanabe, and B.~Xu,
\newblock ``Speaker-conditional chain model for speech separation and extraction,''
\newblock in {\em Proc. Interspeech}, 2020.

\bibitem{delcroix2020improving}
M.~Delcroix, T.~Ochiai, K.~Zmolikova, K.~Kinoshita, N.~Tawara, T.~Nakatani, and S.~Araki,
\newblock ``Improving speaker discrimination of target speech extraction with time-domain {SpeakerBeam},''
\newblock in {\em Proc. ICASSP}, 2020.

\bibitem{smith2005development}
L.~Smith and M.~Gasser,
\newblock ``The development of embodied cognition: Six lessons from babies,''
\newblock {\em Artif. Life}, vol. 11, no. 1--2, pp. 13--29, 2005.

\bibitem{edelman1987neural}
G.~M. Edelman,
\newblock {\em Neural Darwinism: The theory of neuronal group selection.},
\newblock Basic Books, 1987.

\bibitem{ma2009lip}
W.~J. Ma, X.~Zhou, L.~A. Ross, J.~J. Foxe, and L.~C. Parra,
\newblock ``Lip-reading aids word recognition most in moderate noise: a {B}ayesian explanation using high-dimensional feature space,''
\newblock {\em PloS ONE}, vol. 4, no. 3, pp. e4638, 2009.

\bibitem{golumbic2013visual}
E.~Z. Golumbic, G.~B. Cogan, C.~E. Schroeder, and D.~Poeppel,
\newblock ``Visual input enhances selective speech envelope tracking in auditory cortex at a “cocktail party”,''
\newblock {\em J. Neurosci.}, vol. 33, no. 4, pp. 1417--1426, 2013.

\bibitem{crosse2016eye}
M.~J. Crosse, G.~M. Di~Liberto, and E.~C. Lalor,
\newblock ``Eye can hear clearly now: inverse effectiveness in natural audiovisual speech processing relies on long-term crossmodal temporal integration,''
\newblock {\em J. Neurosci.}, vol. 36, no. 38, pp. 9888--9895, 2016.

\bibitem{chung2020facefilter}
S.-W. Chung, S.~Choe, J.~S. Chung, and H.-G. Kang,
\newblock ``{FaceFilter}: Audio-visual speech separation using still images,''
\newblock in {\em Proc. Interspeech}, 2020.

\bibitem{pan2021reentry}
Z.~Pan, R.~Tao, C.~Xu, and H.~Li,
\newblock ``Selective listening by synchronizing speech with lips,''
\newblock {\em IEEE/ACM Trans. Audio, Speech, Lang. Process.}, vol. 30, pp. 1650--1664, 2022.

\bibitem{pan2022seg}
Z.~Pan, X.~Qian, and H.~Li,
\newblock ``Speaker extraction with co-speech gestures cue,''
\newblock {\em IEEE Signal Process. Lett.}, vol. 29, pp. 1467--1471, 2022.

\bibitem{michelsanti2021overview}
D.~Michelsanti, Z.-H. Tan, S.-X. Zhang, Y.~Xu, M.~Yu, D.~Yu, and J.~Jensen,
\newblock ``An overview of deep-learning-based audio-visual speech enhancement and separation,''
\newblock {\em IEEE/ACM Trans. Audio, Speech, Lang. Process.}, 2021.

\bibitem{afouras2018conversation}
T.~Afouras, J.~S. Chung, and A.~Zisserman,
\newblock ``The conversation: Deep audio-visual speech enhancement,''
\newblock in {\em Proc. Interspeech}, 2018.

\bibitem{wu2019time}
J.~{Wu}, Y.~{Xu}, S.~{Zhang}, L.~{Chen}, M.~{Yu}, L.~{Xie}, and D.~{Yu},
\newblock ``Time domain audio visual speech separation,''
\newblock in {\em Proc. ASRU}, 2019.

\bibitem{wu2022time}
Y.~Wu, C.~Li, J.~Bai, Z.~Wu, and Y.~Qian,
\newblock ``Time-domain audio-visual speech separation on low quality videos,''
\newblock in {\em Proc. ICASSP}. IEEE, 2022.

\bibitem{li2020deep}
C.~Li and Y.~Qian,
\newblock ``Deep audio-visual speech separation with attention mechanism,''
\newblock in {\em Proc. ICASSP}, 2020.

\bibitem{li23ja_interspeech}
J.~Li, M.~Ge, Z.~Pan, R.~Cao, L.~Wang, J.~Dang, and S.~Zhang,
\newblock ``Rethinking the visual cues in audio-visual speaker extraction,''
\newblock in {\em Proc. Interspeech}, 2023.

\bibitem{tavcse2022}
J.~Li, M.~Ge, Z.~Pan, L.~Wang, and J.~Dang,
\newblock ``{VCSE}: Time-domain visual-contextual speaker extraction network,''
\newblock in {\em Proc. Interspeech}, 2022.

\bibitem{tan2020audio}
K.~Tan, Y.~Xu, S.-X. Zhang, M.~Yu, and D.~Yu,
\newblock ``Audio-visual speech separation and dereverberation with a two-stage multimodal network,''
\newblock {\em IEEE J. Sel. Topics in Signal Process.}, vol. 14, no. 3, pp. 542--553, 2020.

\bibitem{lu2019audio}
R.~Lu, Z.~Duan, and C.~Zhang,
\newblock ``Audio-visual deep clustering for speech separation,''
\newblock {\em IEEE/ACM Trans. Audio, Speech, Lang. Process.}, vol. 27, no. 11, pp. 1697--1712, 2019.

\bibitem{morrone2019face}
G.~Morrone, S.~Bergamaschi, L.~Pasa, L.~Fadiga, V.~Tikhanoff, and L.~Badino,
\newblock ``Face landmark-based speaker-independent audio-visual speech enhancement in multi-talker environments,''
\newblock in {\em Proc. ICASSP}, 2019.

\bibitem{ephrat2018looking}
A.~Ephrat, I.~Mosseri, O.~Lang, T.~Dekel, K.~Wilson, A.~Hassidim, W.~T. Freeman, and M.~Rubinstein,
\newblock ``Looking to listen at the cocktail party: a speaker-independent audio-visual model for speech separation,''
\newblock {\em ACM Trans. Graph.}, vol. 37, no. 4, pp. 1--11, 2018.

\bibitem{ochiai2019multimodal}
T.~Ochiai, M.~Delcroix, K.~Kinoshita, A.~Ogawa, and T.~Nakatani,
\newblock ``Multimodal {SpeakerBeam}: Single channel target speech extraction with audio-visual speaker clues,''
\newblock in {\em Proc. Interspeech}, 2019.

\bibitem{luo2019conv}
Y.~Luo and N.~Mesgarani,
\newblock ``Conv-{TasNet}: Surpassing ideal time–frequency magnitude masking for speech separation,''
\newblock {\em IEEE/ACM Trans. Audio, Speech, Lang. Process.}, vol. 27, no. 8, pp. 1256--1266, 2019.

\bibitem{pan2020muse}
Z.~Pan, R.~Tao, C.~Xu, and H.~Li,
\newblock ``{MuSE}: Multi-modal target speaker extraction with visual cues,''
\newblock in {\em Proc. ICASSP}, 2021.

\bibitem{pan2023imaginenet}
Z.~Pan, W.~Wang, M.~Borsdorf, and H.~Li,
\newblock ``{ImagineNet}: Target speaker extraction with intermittent visual cue through embedding inpainting,''
\newblock in {\em Proc. ICASSP}, 2023.

\bibitem{luo2020dual}
Y.~Luo, Z.~Chen, and T.~Yoshioka,
\newblock ``Dual-path {RNN}: Efficient long sequence modeling for time-domain single-channel speech separation,''
\newblock in {\em Proc. ICASSP}, 2020.

\bibitem{usev21}
Z.~Pan, M.~Ge, and H.~Li,
\newblock ``{USEV}: Universal speaker extraction with visual cue,''
\newblock {\em IEEE/ACM Trans. Audio, Speech, Lang. Process.}, vol. 30, pp. 3032--3045, 2022.

\bibitem{blanco2023avse}
A.~L.~A. Blanco, C.~Valentini-Botinhao, O.~Klejch, M.~Gogate, K.~Dashtipour, A.~Hussain, and P.~Bell,
\newblock ``{AVSE} challenge: Audio-visual speech enhancement challenge,''
\newblock in {\em Proc. SLT}, 2023.

\bibitem{pan2022hybrid}
Z.~Pan, M.~Ge, and H.~Li,
\newblock ``A hybrid continuity loss to reduce over-suppression for time-domain target speaker extraction,''
\newblock in {\em Proc. Interspeech}, 2022.

\bibitem{wang2023tf}
Z.-Q. Wang, S.~Cornell, S.~Choi, Y.~Lee, B.-Y. Kim, and S.~Watanabe,
\newblock ``{TF-GridNet}: Making time-frequency domain models great again for monaural speaker separation,''
\newblock in {\em Proc. ICASSP}, 2023.

\bibitem{Cornell2022}
S.~Cornell, Z.-Q. Wang, Y.~Masuyama, S.~Watanabe, M.~Pariente, and N.~Ono,
\newblock ``Multi-channel target speaker extraction with refinement: {T}he {WAVL}ab submission to the second clarity enhancement challenge,''
\newblock in {\em Proc. Clarity}, 2022.

\bibitem{Afouras18b}
T.~Afouras, J.~S. Chung, and A.~Zisserman,
\newblock ``Deep lip reading: A comparison of models and an online application,''
\newblock in {\em Proc. Interspeech}, 2018.

\bibitem{le2019sdr}
J.~Le~Roux, S.~Wisdom, H.~Erdogan, and J.~R. Hershey,
\newblock ``{SDR}--half-baked or well done?,''
\newblock in {\em Proc. ICASSP}, 2019.

\bibitem{afouras2018lrs3}
T.~Afouras, J.~S. Chung, and A.~Zisserman,
\newblock ``{LRS3-TED}: a large-scale dataset for visual speech recognition,''
\newblock {\em arXiv preprint arXiv:1809.00496}, 2018.

\bibitem{graetzer2021clarity}
S.~Graetzer, J.~Barker, T.~J. Cox, M.~Akeroyd, J.~F. Culling, G.~Naylor, E.~Porter, and R.~V. Munoz,
\newblock ``Clarity-2021 challenges: Machine learning challenges for advancing hearing aid processing,''
\newblock in {\em Proc. Interspeech}, 2021, vol.~2.

\bibitem{thiemann2013demand}
J.~Thiemann, N.~Ito, and E.~Vincent,
\newblock ``{DEMAND}: a collection of multi-channel recordings of acoustic noise in diverse environments,''
\newblock in {\em Proc. Meetings Acoust.}, 2013.

\bibitem{dubey2023icassp}
H.~Dubey, A.~Aazami, V.~Gopal, B.~Naderi, S.~Braun, R.~Cutler, H.~Gamper, M.~Golestaneh, and R.~Aichner,
\newblock ``Deep speech enhancement challenge at {ICASSP} 2023,''
\newblock in {\em Proc. ICASSP}, 2023.

\bibitem{fonseca2021fsd50k}
E.~Fonseca, X.~Favory, J.~Pons, F.~Font, and X.~Serra,
\newblock ``{FSD50K}: An open dataset of human-labeled sound events,''
\newblock {\em IEEE/ACM Trans. Audio, Speech, Lang. Process.}, vol. 30, pp. 829--852, 2021.

\bibitem{reddy2021dnsmos}
C.~Reddy, V.~Gopal, and R.~Cutler,
\newblock ``{DNSMOS}: A non-intrusive perceptual objective speech quality metric to evaluate noise suppressors,''
\newblock in {\em Proc. ICASSP}, 2021.

\end{thebibliography}

\end{document}